\documentclass{amsart}
\usepackage{amsmath, amssymb, amsthm, epsfig, rotating}
\usepackage{graphicx}

\newtheorem{thm}{Theorem}
\newtheorem{prop}{Proposition}

\newtheorem{cor}{Corollary}
\newcommand{\pf}{{\noindent\bf Proof:\,}}

\begin{document}
\title{The maximum number of minimal codewords in an $[n,k]-$code }
\author{A. Alahmadi}
\address{MECAA, Math Dept of King Abdulaziz University, Jeddah, Saudi Arabia }
\email{adelnife2@yahoo.com}
\author{R.E.L. Aldred}
\address{Department of Mathematics and Statistics,
University of Otago,
P. O. Box 56, Dunedin, New Zealand}
\email{raldred@maths.otago.ac.nz}
\author{R. dela Cruz}
\address{Division of Mathematical Sciences, SPMS, Nanyang Technological University, Singapore
\\and\\Institute of Mathematics, University of the Philippines Diliman, Quezon City, Philippines}
\email{ROMA0001@e.ntu.edu.sg}
\author{P. Sol\'e}
\address{Telecom ParisTech, 46 rue Barrault, 75634
Paris Cedex 13, France.\\and\\MECAA, Math Dept of King Abdulaziz University, Jeddah, Saudi Arabia }
\email{sole@enst.fr}
\author{C. Thomassen}
\address{Department of Mathematics,
Technical University of Denmark,
DK-2800 Lyngby, Denmark\\and\\MECAA, Math Dept of King Abdulaziz University, Jeddah, Saudi Arabia }
\email{C.Thomassen@mat.dt.u.dk}
\begin{abstract}
Upper and lower bounds are derived for the quantity in the title, which is tabulated
for modest values of $n$ and $k.$ An application to graphs with many cycles is given.
\end{abstract}

\maketitle
\section{Introduction}
Consider a binary linear code $C.$ A codeword of $C$ is called {\bf minimal} is its support does not contain properly the support of another nonzero codeword. This concept was discovered independently in code-based secret sharing schemes \cite{AB} and also in the study of the Voronoi domain of a code in the context of decoding \cite{A}.
What is the maximum number $M(C)$ of minimal codewords a code $C$ of given length and dimension might have?
 Formally, denoting by ${\bf  C}[n,k]$
the set of all $[n,k]$ codes, we define, following the companion paper \cite{AADST}, the function
$$M(n,k)=\max \{M(C) :\, C \in {\bf C}[n,k]\},$$ as the maximum of $M(C)$ over that set of codes.
While the concern of  \cite{AADST} was asymptotics, we will consider in this note only bounds on
or exact values of that function for finite values of $n$ and $k.$
We will consider three upper bounds. The so-called trivial bound, the matroid bound  as in \cite{AADST} and the Agrell bound \cite{A}. We derive a recursive inequality on $M(n,k)$ which gives an alternative proof of the matroid bound, independent of matroid theory as a special case.  The connection with intersecting codes shows that the trivial bound is sharp when $k$ is small compared to $n.$
The Agrell bound which is asymptotically equivalent to the trivial bound can be sharper for finite values of $n$ and $k$ when $k$ is close to $n.$
 In particular in the special case of the cycle code of graphs this bound is a sharpening of the $\frac{15}{16}$ bound of \cite{AT} in a special case. For lower bounds, neither the random coding bound of \cite{AB} nor the combinatorial bound of \cite{ABCH} matches
 explicit constructions. 

The material is organized as follows. Section 2 is dedicated to upper bounds. Section 3 considers lower bounds. Section 4 builds a table of values of  and bounds on $M(n,k)$ for $k \le 13$
and $ n\le 15.$

\section{Upper  bounds}
An immediate upper bound is $M(n,k)\le 2^k-1.$ We call this the trivial bound.
Another upper bound derived in \cite{AADST} by use of matroid theory  is
$$M(C)\le { n \choose k-1} ,$$
which is sharper than the trivial bound at high rates. We give a recurrence relation that implies the matroid bound.

{\thm For all $1\le k\le n$ we have $$ M(n,k)\le M(n-1,k-1)+{n-1\choose k-1}. $$}

\pf
Let $H$ be the parity check of $C$ that realizes $M(n,k),$ and $H'$ the matrix with column $n$
removed. Assume, up to column reordering,  that there is a basis of the column space not containing column $n,$ or equivalently
that the rank of $H'$ is $n-k.$ This is always possible if $k\ge 1.$ Let $x$ be a nonzero minimal codeword  in $C$ and discuss according to the value
 of $x_n.$\\

If $x_n=0$ then the projection $x'$  on the first $n-1$ coordinates is a minimal codeword in $Ker(H')$
an $[n-1,k-1]$ code. Therefore there are at most $M(n-1,k-1)$ such vectors. \\

If $x_n=1$ then the set of columns  where the projection $x'$  on the first $n-1$ coordinates is nonzero
form an independent  set of the column space
of $H',$ because of the minimality property of $x.$ There are at most

$${n-1\choose n- k}={n-1\choose k-1}$$

 possible such $x'.$
\qed

The matroid bound now follows as a Corollary of the above Theorem.

{\cor For all $1\le k\le n$ we have $M(n,k)\le {n\choose k-1}.$ }

\pf We reason
by induction on  $k.$ Clearly, the bound
is true for $k=1,$ since $M(n,1)=1={n\choose 0}.$  Assuming $M(n-1,k-1)\le {n-1\choose k-2},$
by Pascal's triangle, using the above theorem, we are done.
\qed

Another upper bound is given in \cite[Theorem 5]{A}.

{\thm For $\frac{k-1}{n} > \frac{1}{2}$ we have $$ M(n,k)  \le \frac{2^k}{4n(\frac{k-1}{n}-\frac{1}{2})^2}.$$ }

A difficult problem in graph theory is to bound above the maximum number of cycles a connected graph on  $p$ vertices and with $q$ edges can have \cite{ES}. The analogue of the trivial bound in that context is $2^{q-p+1}.$  The first bound significantly below that was \cite{AT}. The next result is a strengthening for graphs of average degree $>4$ of  that result .

{\cor  If $\Gamma$ is a connected graph on $p$ vertices and with $q$ edges satisfying $q>2p$ then its number of elementary cycles is at most
$$ \frac{q2^{q-p+1}}{(q-2p)^2}  .$$
}

\pf Recall that with every connected graph $\Gamma$ on $p$ vertices and with $q$ edges is attached a binary $[q, q-p+1]$ code $C(\Gamma)$ called the cycle
code of the graph. Its codewords are indicator vectors of either elementary cycles or edge disjoint unions of such. The minimal codewords of  $C(\Gamma)$ are the indicator vectors of the (elementary) cycles of the graph.
The result follows by applying Theorem 2 to that code, after some algebra.
\qed

The bound in \cite{AT} was $ \frac{15}{16} 2^{q-p+1}  .$ The last result is  sharper for $p\ge 1$ as can be seen by computing the discriminant of a quadratic equation in $q.$
\section{Lower bounds}

As in  \cite{AADST}  there is a random coding lower bound from
\cite{AB}.

$$M(n,k)2^{n-k} \ge \sum_{j=0}^{n-k+1}{n \choose j}\prod_{i=0}^{j-2}(1-2^{-(n-k-i)}).$$
Another existence bound is as follows.
Denote by $d(n,k)$ the largest minimum distance of an $[n,k]$ code. The following Proposition is a direct consequence of \cite[Prop. 2.1.]{ABCH}.

{\prop For all $n\ge k\ge 1,$ we have
$$  \sum_{i=1}^{\lfloor n/d(n,k)\rfloor}{M(n,k)\choose i} \ge 2^k-1.$$
 }

\pf 

Let $C$ be an $[n,k,d]$ code. By induction it can be seen that every nonzero codeword can be written as a sum of at most $\lfloor n/d\rfloor$ support disjoint minimal codewords.
Hence, enumeration of such sums yields
$$  \sum_{i=1}^{\lfloor n/d\rfloor}{M(C)\choose i} \ge 2^k-1.$$
The result follows by choosing $C$ to be optimal for $d.$
\qed

This result shows that good codes cannot have too few minimum codewords. It is not very sharp. We only get $M(8,4) \ge 5,$ when the example of the extended Hamming code shows that $M(8,4) \ge 14.$ 
\section{Tabulating $M(n,k).$}
\subsection{Monotonicity properties}
It is easy to show that $M(n,k)\le M(n+1,k)$ by adding a zero column to a code realizing $M(n,k).$ This innocent remark provides better bounds that the random coding bound for
$k=4$ and $n=7,8$ as well as $k=5$ and $n=7,8,9$ and so on.
It is false that $M(n,k)\le M(n,k+1)$ as the values of $M(4,k)$ already show. We conjecture, but cannot prove,  that $M(n,k)$ is an unimodal function of $k$ for fixed $n.$
{\prop \label{decomp} For binary codes $C,D$ we have  $M(C\oplus D)=M(C)+M(D).$ }

\pf 
If $c$ and $d$ are minimal codewords of respectively $C$ and $D$ then $(c,0)$ and $(0,d)$ are minimal codewords of $C\oplus D.$ Conversely, we claim that all minimal codewords of the latter code arise in that way.Indeed if $(c,d)$ is a minimal  codeword of  $C\oplus D,$ with both $c$ and $d$ nonzero, then $(c,d)=(c,0)+(0,d)$ contradicting minimality.

\qed

Using the above Proposition,
we see that $M(n,k)$  is super additive
$$M(n+m,k+j)\ge M(n,k)+M(m,j).$$

\subsection{Exact values}
Trivial values are $M(n,1)=1,$ and $M(n,n)=n$ for all $n\ge 1.$ Already $M(n,n-1)$ is known but
requires a proof.
{\prop  $M(n,n-1)= {n \choose 2}$ for $ 3\le n .$}

\pf
We claim that
$$M(n,n-1)=\max\{ {x \choose 2}+(n-x)| 2\le x \le n\}.$$
Indeed, denoting by $P_x$ the parity-check code of length $x$ and by $U_y$ the universe code of length $y,$ we see that, by Proposition \ref{decomp},  $M(P_x\oplus U_{n-x})= {x \choose 2}+(n-x)=:f(x)$
Studying the variation of the quadratic $f(x)$ shows that it is increasing for $x\in[2,n]$. Since $f(n)={n \choose 2},$ we are done.
\qed
\subsection{Intersecting codes}
Recall that a code is {\bf  intersecting }\cite{S}  if the respective supports of any two nonzero codewords intersect.
As observed in \cite{AADST} a linear binary code meets the trivial bound with equality iff it is intersecting. Following \cite{S}, denote by $f(k)$ the shortest length of a binary linear intersecting code. Equivalently, there is a function $g(n)$ such that if $k\le g(n)$
 then there is an intersecting $[n,k]$ code; and there is not if $k>n;$
thus if $k\le g( n)$ then $M(n,k)=2^k-1,$ and  if $k> g( n)$ then $M(n,k)\le 2^k-2.$ The function $g(n),$ the inverse of $f,$ is known exactly for $1\le n\le 15$ \cite{S} and given in Table 1.
\begin{table}[ht]
\begin{center}
\small
\begin{tabular}{|c|c|c|c|c|c|c|c|c|c|c|c|c|c|}
\hline
$n$ &3&4&5&6&7&8&9& 10&11&12&13&14&15\\
\hline
$g(n)$ &2&2&2&3&3&3&4&4&4&4&5&5&6\\
\hline
\end{tabular}
\label{table1}
\caption{$g(n)$ for $3\le n\le 15$}
\end{center}
\end{table}
\subsection{Table of  $M(n,k)$}

The exponents in Table 2 are as follows.
\begin{itemize}
 \item t - Trivial bound
 \item m - Matroid bound
 \item a - Agrell bound
\end{itemize}
When the trivial bound is met with equality the exponent $t$ is omitted. Empty entries correspond to $k>n$ when $M(n,k)$ is undefined.
The lower bounds are derived by explicit constructions of codes $C$  derived from A. Betten list of indecomposable codes \cite{B}, followed  by application of rule G or rule H of \cite{A} to derive $M(C).$
The codes realizing $M(n,k)$ are not in general optimal for the minimum distance, but they are in general indecomposable.
\begin{table}

\scriptsize
\begin{tabular}{|l|l|l|l|l|l|l|l|l|l|l|l|l|l|}\hline
$n/k$ &	1 & 2 &	3 & 4	           	& 5	             & 6		  & 7	                 & 8			& 9	                & 10	                & 11	                & 12	                 & 13	                \\\hline
3   &	1 & 3 &	3 &		        &		     &			  &		         &			&		        &		        &		        &		         &	 	        	\\\hline
4   &	1 & 3 &	6 &      4	        &		     &			  &		         &			&		        &		        &		        &		         &		        	\\\hline
5   &	1 & 3 &	6 &	10	        & 5		     & 		          &		         &			&		        &		        &		        &		         &		        	\\\hline
6   &	1 & 3 &	7 & 11-$14^{\text{t}}$	& 15                 & 6	          &		         &			&		        &		        &		        &		         &		       \\\hline
7   &	1 & 3 &	7 &	14	        & 17-$30^{\text{t}}$ & 21		  & 7		         &			&		        &		        &		        &		         &		        \\\hline
8   &	1 & 3 &	7 &	14	        & 22-$30^{\text{t}}$ & 25-$55^{\text{m}}$ & 28		         & 8			&		        &		        &		        &		         &		        	\\\hline
9   &	1 & 3 &	7 &	15	        & 26-$30^{\text{t}}$ & 33-$62^{\text{t}}$ & 36-$83^{\text{m}}$	 & 36	        & 9		        &		        &		        &		         &		        	\\\hline
10  &	1 & 3 &	7 &	15		& 30		     & 42-$62^{\text{t}}$ & 48-$126^{\text{t}}$	 & 48-$119^{\text{m}}$	& 45	        & 10		        & 		        &		         &		        	\\\hline
11  &	1 & 3 &	7 &	15		& 30		     & 52-$62^{\text{t}}$ & 66-$126^{\text{t}}$	 & 69-$254^{\text{t}}$	& 63-$164^{\text{m}}$  & 55		        & 11		        &		         &		        	\\\hline
12  &	1 & 3 &	7 &	15		& 30	             & 54-$62^{\text{t}}$ & 90-$126^{\text{t}}$	 & 103-$254^{\text{t}}$	& 95-$384^{\text{a}}$  & 82-$219^{\text{m}}$   & 66		        & 12		         &		        	\\\hline
13  &	1 & 3 &	7 &	15		& 31	             & 58-$62^{\text{t}}$ & 94-$126^{\text{t}}$	 & 151-$254^{\text{t}}$	& 149-$510^{\text{t}}$  & 130-$532^{\text{a}}$  & 102-$285^{\text{m}}$  & 78	         & 13		        	\\\hline
14  &	1 & 3 &	7 &	15		& 31	             & 62                 & 106-$126^{\text{t}}$ & 159-$254^{\text{t}}$	& 245-$510^{\text{t}}$  & 217-$896^{\text{a}}$  & 175-$796^{\text{a}}$  & 126-$363^{\text{m}}$   & 91	        \\\hline
15  &	1 & 3 &	7 &	15		& 31	             & 63	          & 108-$126^{\text{t}}$ & 171-$254^{\text{t}}$ & 245-$510^{\text{t}}$  & 385-$1022^{\text{t}}$ & 308-$1228^{\text{a}}$ & 221-$1253^{\text{a}}$ & 155-$454^{\text{m}}$ \\
\hline\end{tabular}

\caption{$M(n,k)$ for $3\leq n\leq 15$ and $k\le 13$}
\end{table}
\section{Conclusion and open problems}
In this note we have considered the function $M(n,k)$  maximum number of minimal codewords of a binary linear code of parameters $[n,k].$ Three  upper bounds have been considered in turn: trivial, Agrell and matroid bound. From Table 2 we see that they are most relevant respectively at low rate, high rate and very high rate. Lower bounds have been
derived by selecting suitable  indecomposable codes. It seems possible but computationally heavy to derive the exact values of $M(n,k)$ by combining Proposition \ref{decomp} with a database of indecomposable codes that would be more comprehensive than that of \cite{B} where only indecomposable codes with large minimum distance are listed.
We conjecture that the lower bounds in Table 2 are in fact exact values, and that they are obtained for indecomposable codes.

{\bf Acknowledgement:} The authors are grateful to Alfred Wassermann for providing reference  \cite{B}.


\begin{thebibliography}{99}
\bibitem{A}E. Agrell, On the Voronoi neighbor ratio for binary linear codes, IEEE Transactions in Information Theory (1998) 3064--3072.
\bibitem{ABCH}A. Ashikmin, A.  Barg, G. Cohen, L. Huguet, Variations on minimimal codewords in linear codes, Springer LNCS 948 (1995) 96-–105.
\bibitem{AT} R.E.L. Aldred. C. Thomassen  On the maximum number of cycles in a planar graph, J. Graph Theory {\bf 53} (2008) 255--264.

\bibitem{AADST}A. Alamadhi, R.E.L. Aldred, R. de la Cruz, P. Sol\'e, C. Thomassen, The maximum number of minimal codewords in long codes, submitted.
\bibitem{B} {\tt http://www.math.colostate.edu/\~\,betten/research/codes/GF2/codes\_GF2.html}
\bibitem{AB}A. Ashikmin, A.  Barg, Minimal vectors in linear codes, IEEE Transactions in Information Theory (1998) 2010--2017.
\bibitem{DSL}Gy.  Dosa, I.  Szalkai, C.  Laflamme, The maximum and minimum number of circuits and bases of matroids, PU. M. A. {\bf 15} (2004) 383--392.
\bibitem{ES}R.C.  Entringer, P.J.  Slater, On the maximum number of cycles in a graph, Ars Combinatoria {\bf 11} (1981) 289--294. 
\bibitem{S}  N. J. A. Sloane, Covering Arrays and Intersecting Codes, J. Combinatorial Designs, 1 (1993), 51--63.
\end{thebibliography}
\end{document}